\documentclass[pra, twocolumn, showpacs, showkeys]{revtex4-1}

\usepackage{amsmath}
\usepackage{amssymb}
\usepackage{mathrsfs}
\usepackage[dvips]{graphicx}
\usepackage{subfigure}
\usepackage{array}

\DeclareGraphicsExtensions{.ps}

\newcommand*{\etal}{\textit{et al.}}
\def\H1m{H^{\scriptscriptstyle (1)}_m}
\def\leta{\tilde{\lambda}^{a,n}}
\def\letb{\tilde{\lambda}^{b,n+1}}

\def\ledb{\tilde{\lambda}^{b,n+1}(\theta)}

\begin{document}
\title{Cold atom scattering by cavity fields in a two-dimensional geometry}
\date{\today}
\author{John Martin}
\email{jmartin@ulg.ac.be}
\author{Thierry Bastin}
 \affiliation{Institut de Physique Nucl\'eaire, Atomique et de
Spectroscopie, Universit\'e de Li\`ege, 4000
Li\`ege, Belgium}

\begin{abstract}
The quantum theory of the cold atom scattering by cavity fields in a two-dimensional geometry is presented.
A distinct regime from the usual Raman-Nath, Bragg and Stern-Gerlach regimes is investigated, considering the situation where the cavity light field acts as a repulsive and an attractive two-dimensional potential. General expressions for the scattering lengths (the two-dimensional analogues to the three-dimensional scattering cross-sections) of
finding the atoms deexcited or not after their interaction with the cavity are derived. The connection with the classical Rabi limit when the incident atomic kinetic energy is high compared with the atom-field interaction energy is made. In the cold atom regime characterized by much lower incident atomic kinetic energies, the scattering process exhibits very peculiar properties in connection with quasibound states of the atomic motion induced by the attractive potential of the cavity light field.
\end{abstract}



\maketitle

\section{Introduction}
\label{Introduction}


Scattering of particles by light has been the focus of many investigations since it was predicted by Kapitza and Dirac that a standing wave of light can diffract electrons~\cite{Kap33}. Several decades thereafter it was suggested that neutral atoms can experience much stronger diffraction effects by light fields~\cite{Afb66, Ash70}. Since then, many experimental observations of those effects were reported using optical gratings formed by optical monochromatic standing waves (see, e.g., Ref.~\cite{Mey01} and references therein). Three major regimes are usually distinguished~\cite{Mey01}~: the Raman-Nath, Bragg, and Stern-Gerlach regimes. In the first two cases, the width of the incident atomic beam is large in comparison with the period of the standing wave pattern. The latter case considers the inverse situation where this width is much smaller than this period. In that case, it is predicted that the light field gradients can produce state-selective deflection of an atomic beam in the plane formed by the standing wave and the atomic beam propagation axes. The first experimental verification of the optical Stern-Gerlach effect was reported by Sleator \textit{et al.}~\cite{Sle92}. There the atoms are considered fast enough so that their motion along their propagation axis can be considered as classical. In contrast the component of velocity along the standing wave axis is treated quantum mechanically and the Stern-Gerlach effect stems from the interplay between that motion, the internal degrees of freedom, and the light field gradients in the optical grating.

In this paper we are interested in considering a regime where the atoms sent through the standing light wave are rather very slow with a kinetic energy lower or of the same order of magnitude as their interaction energy with the light field. In that case the motion of the atoms along their propagation axis must be treated quantum mechanically. For one dimensional problems describing moving atoms in interaction with cavity light fields, it was shown that this regime gives rise to heaps of new phenomena~\cite{Eng91, *Scu96, *Mey97, *Lof97, *Bas03b, *Mar07}. We will focus on the situation where the wavepacket extension in the direction along the stationary light field axis is much smaller than the light wavelength. We consider further that the atoms are sent at an extremum of the standing light wave so as to avoid the Stern-Gerlach effect. This allows us not to consider the longitudinal motion along the standing light wave and to focus on the significant diffraction effects of the atomic beam in the transverse plane that are to be observed owing to the slowness of the incident atoms. The theoretical developments are made in the framework of cavity QED for which the atoms interact with the waist of a quantized field mode of a cavity. Throughout the paper we will be interested in symmetrical cavity mode functions (like the fundamental mode of a Fabry-Perot cavity) for which the powerful method of partial waves and phase shifts in two dimensions can be used. The cavity mode frequency is supposed to be equal to a transition of the incident atoms considered as two-level atoms.

The paper is organized as follows. In Sec.~\ref{ModelSection}, the Hamiltonian and wave functions of the system under consideration are presented using the formalism of the two-dimensional scattering theory. We further derive the state-dependent differential and total scattering lengths, that are in a two-dimensional geometry what the scattering cross-sections are in three dimensions. These scattering lengths are then analyzed in Sec.~\ref{TSLSection} in two different regimes defined as a function of the incident atomic kinetic energy compared with the atom-field interaction energy. We finally draw conclusions in Sec.~\ref{SummarySection}.

\section{Model}
\label{ModelSection}

\begin{figure}
\begin{center}
\includegraphics[width=.95\linewidth]{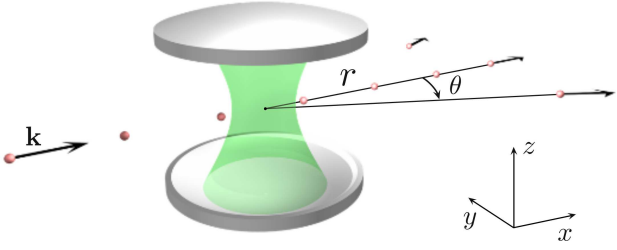}
\end{center}
\caption{(Color online) Geometry of the investigated scattering problem. Incident atoms with wave vector $\mathbf{k}$ are sent towards the beam waist of an open cavity and are scattered in the transverse $x-y$ plane. The $x$ axis is defined along the incident beam direction and the $z$ direction is along the cavity axis. In the plane of motion, $r$ and $\theta$ define the polar coordinates of the particles with the origin set at the cavity center.} \label{fig1}
\end{figure}

We consider a two-level atom moving along the $x$ direction on the way to a cavity. The atom is coupled resonantly to a single mode of the quantized cavity field. The atomic center-of-mass motion is restricted to the $x-y$ plane (see Fig.~\ref{fig1}). This motion is described quantum mechanically and the usual rotating-wave approximation is made. In the interaction picture, the atom-field Hamiltonian reads
\begin{equation}
    \label{Hamiltonian}
        H = \frac{\mathbf{p}^2}{2M}
        + \hbar g \, v(\mathbf{r}) (a^{\dagger} \sigma + a
        \sigma^{\dagger}),
\end{equation}
where $M$ is the atomic mass, $\mathbf{r}$ and $\mathbf{p}$ are, respectively, the atomic center-of-mass position and momentum in the $x-y$ plane, $\sigma$ is the projection operator $|b\rangle \langle a|$, with $|a\rangle$ [$|b\rangle$] the upper [lower] level of the atomic transition, $a$ and $a^{\dagger}$ are, respectively, the annihilation and creation operators of the cavity radiation field, $g$ is the atom-field coupling strength, and $v(\mathbf{r})$ is the cavity field mode function. Hereafter the global atom-field state is denoted by $|\psi(t)\rangle$ and the cavity field eigenstates by $|n\rangle$.

The incoming atom is described by a plane wave with wave vector $\mathbf{k}$ (monokinetic atom). It is supposed to be initially in the excited state $|a\rangle$ and the cavity field in the Fock state $|n\rangle$. The generalization to other initial states would proceed along the same lines. The initial atom-field wave function reads
\begin{equation}\label{decompdd}
    \langle \mathbf{r}|\psi(0)\rangle= e^{ikx}|a,n\rangle = \frac{e^{ikx}}{\sqrt{2}}\left(\hspace{0.5pt}|+_n\rangle+|-_n\rangle\right),
\end{equation}
where we have used the usual dressed-state basis vectors
\begin{equation}\label{basis}
|\pm_{n}\rangle = \frac{1}{\sqrt{2}}\left(|a,n\rangle \pm |b,n+1\rangle\right).
\end{equation}

In that case the wave function components
\begin{equation}
\psi^{\pm_n}(\mathbf{r},t)=\langle \mathbf{r},\pm_n|\psi(t)\rangle
\end{equation}
read initially $\psi^{\pm_n}(\mathbf{r},0)=e^{ikx}/\sqrt{2}$ and obey the Schr\"odinger equation
\begin{equation}\label{eqschro2dtime2}
i\hbar \frac{\partial}{\partial t}\psi^{\pm_n}(\mathbf{r},t)
=\left[-\frac{\hbar^2}{2M}\nabla^2+V^{\pm_n}(\mathbf{r})\right]\psi^{\pm_n}(\mathbf{r},t),
\end{equation}
with
\begin{equation}
    V^{\pm_n}(\mathbf{r})= \pm\hbar g\sqrt{n+1}\,v(\mathbf{r}).
\end{equation}


Equation~(\ref{eqschro2dtime2}) expresses that the atom-field interaction reduces to a \emph{two-dimensional} scattering problem where each $\psi^{\pm_n}(\mathbf{r},t)$ wave function component is subjected to the potential $V^{\pm_n}(\mathbf{r})$.

Setting
\begin{equation}
    |\psi(t)\rangle = e^{-iE_kt/\hbar}\,|\varphi\rangle
\end{equation}
with $E_k=\hbar^2k^2/2M$, the time-independent Schr\"{o}dinger equation reads in polar coordinates $(r,\theta)$
\begin{equation}\label{Schropm2dpol}
    \left[\frac{\partial^2}{\partial r^2}+\frac{1}{r}\frac{\partial}{\partial
    r}+\frac{1}{r^2}\frac{\partial^2}{\partial
    \theta^2}+k^2\mp\kappa^2_n\:v(r,\theta)\right]\varphi^{\pm_n}(r,\theta)=0,
\end{equation}
with $\varphi^{\pm_n}(\mathbf{r}) \equiv \langle \mathbf{r}, \pm_n | \varphi \rangle$ and
\begin{equation}
\kappa^2_n=\kappa^2\sqrt{n+1}, \quad \kappa^2 = \frac{2M}{\hbar}\,g.
\end{equation}

Here we will consider mode functions having a cylindrical symmetry, i.e., $v(r,\theta) \equiv v(r)$. In that case the most general solution to Eq.~(\ref{Schropm2dpol}) that is symmetric with respect to the $x$-direction of the incoming particles is given by~\cite{Mor53, *Lap82, *Lap86}
\begin{equation}
\label{solgenpm2dpol}
    \varphi^{\pm_n}(r,\theta)=\sum_{m=0}^{+\infty}R^{\pm_n}_{m}(r)\cos(m\theta),
\end{equation}
where $R^{\pm_n}_{m}(r)$ is the most general solution to the radial equation
\begin{equation}\label{eqradpm}
    \frac{d^2 R^{\pm_n}_{m}}{d r^2}+\frac{1}{r}\frac{d R^{\pm_n}_{m}}{d r}+
    \left(k^2\mp\kappa^2_n\:v(r)-\frac{m^2}{r^2}\right)R^{\pm_n}_{m}=0.
\end{equation}

By setting $u^{\pm_n}_m(r)=\sqrt{r}\,R^{\pm_n}_m(r)$, Eq.~(\ref{eqradpm}) turns into
\begin{equation}\label{eqnradm2d}
    \frac{d^2u^{\pm_n}_m}{dr^2}+\left(k^2\mp\kappa^2_n\:v(r)-\frac{m^2-1/4}{r^2}\right)u^{\pm_n}_m=0.
\end{equation}
For $m>0$ the term $(m^2-1/4)/r^2$ is a centrifugal barrier that prevents a classical particle of energy $\hbar^2k^2/2M$ and angular momentum $m\hbar$ from coming closer to the origin than the critical distance $r_m$ given by
\begin{equation}\label{relle}
    k r_m=\sqrt{m^2-1/4}.
\end{equation}

It is useful to express the stationary wave function components (\ref{solgenpm2dpol}) in a form having the asymptotic behavior
\begin{equation}
    \label{asymptbehavior}
   \varphi^{\pm_n}(r,\theta) \xrightarrow[r\rightarrow \,\infty]{} e^{i k x} + f^{\pm_n}(\theta) \frac{e^{i k r}}{\sqrt{r}},
\end{equation}
so as to identify the differential scattering lengths~\footnote{Scattering lengths are in a two-dimensional geometry what the scattering cross-sections are in three dimensions~\cite{Mor53}.}
\begin{equation}
\lambda^{\pm_n}(\theta) \equiv |f^{\pm_n}(\theta)|^2.
\end{equation}
For that purpose we divide the transverse plane into two regions separated by a circle of radius $R$ in such a way that the mode function $v(r)$ vanishes entirely for $r > R$. The region inside the circle is called the \emph{inside} of the cavity (region I) and the other one the \emph{outside} (region II). Outside the cavity, the solution
(\ref{solgenpm2dpol}) can be written as
\begin{equation}\label{psiBarrierVpmout}
\begin{aligned}
        \varphi^{\pm_n}(r,\theta)={}&\frac{1}{\sqrt{2}}\sum_{m=0}^{+\infty}\epsilon_mi^m\cos(m\theta)J_m(k
        r)\\
        &+\frac{1}{\sqrt{2}}\sum_{m=0}^{+\infty}\epsilon_mB^{\pm_n}_m\cos(m\theta)\H1m(kr),
\end{aligned}
\end{equation}
where
\begin{equation}
\epsilon_m = 2 \,\,\, (m > 0), \quad \epsilon_0 = 1,
\end{equation}
$J_m$ is the Bessel function of the first kind of order $m$, $\H1m$ is the
first Hankel function of order $m$~\cite{Abr70}, and $B^{\pm_n}_m$ are coefficients
determined from the continuity conditions of the wave functions and their derivatives at the interface between regions I and II (at $r = R$).
Equation~(\ref{psiBarrierVpmout}) behaves asymptotically as Eq.~(\ref{asymptbehavior})~: the first term on the right-hand side
is just the incoming plane wave $e^{ikx}/\sqrt{2}$ expanded in polar coordinates in terms of Bessel functions, whereas the second term
represents outgoing cylindrical waves.

Owing to the central symmetry of the cavity mode functions considered here, the total scattering lengths
\begin{equation}
\label{lambdapmn}
\lambda^{\pm_n} \equiv \int_{-\pi}^{\pi} \lambda^{\pm_n}(\theta) d\theta
\end{equation}
may be written in the form~\cite{Mor53}
\begin{equation}
\label{lambdapmnsum}
\lambda^{\pm_n} = \frac{4}{k} \sum_{m = 0}^{\infty} \epsilon_m \sin^2 \delta_m^{\pm_n},
\end{equation}
where the sine arguments $\delta_m^{\pm_n}$ represent half of the phase shifts induced by the scattering potentials $V^{\pm_n}(r)$ in the presence of the centrifugal barrier between an ingoing and the corresponding outgoing scattered cylindrical waves~\cite{Mor53}. The $B^{\pm_n}_m$ coefficients of Eq.~(\ref{psiBarrierVpmout}) read in terms of these phase shifts
\begin{equation}\label{Bps}
    B^{\pm_n}_m=\frac{i^m}{2}\left(e^{2i\delta_m^{\pm_n}}-1\right).
\end{equation}

For many purposes it is interesting to express everything in the $|\mathbf{r},\gamma_n\rangle \equiv$ $\{|\mathbf{r},a,n\rangle, |\mathbf{r},b,n+1\rangle\}$ representation. The wave function components $\varphi^{\gamma_n}(\mathbf{r})$ $\equiv \langle \mathbf{r},\gamma_n|\varphi\rangle$
are simply given by [see Eq.~(\ref{basis})]
\begin{align}
 \varphi^{a,n}(r,\theta) & = \frac{1}{\sqrt{2}}\left( \varphi^{+_n}(r,\theta) + \varphi^{-_n}(r,\theta)\right), \nonumber \\
 \varphi^{b,n+1}(r,\theta) & = \frac{1}{\sqrt{2}}\left( \varphi^{+_n}(r,\theta) - \varphi^{-_n}(r,\theta)\right),
\end{align}
and read here, considering Eq.~(\ref{psiBarrierVpmout}),
\begin{align}
 \varphi^{a,n}(r,\theta) & = e^{ikr\cos\theta} +
                    \sum_{m=0}^{+\infty}\epsilon_mB^{a,n}_m\cos(m\theta)\H1m(kr), \nonumber \\
 \label{compb2d} \varphi^{b,n+1}(r,\theta) & = \sum_{m=0}^{+\infty}\epsilon_mB^{b,n+1}_m\cos(m\theta)\H1m(kr),
\end{align}
with
\begin{align}
B^{a,n}_m & \equiv \frac{B^{+_n}_m+B^{-_n}_m}{2} = \frac{i^m}{4}\left(e^{2i\delta_m^{+_n}}+e^{2i\delta_m^{-_n}}-2\right), \nonumber \\
\label{Bab} B^{b,n+1}_m & \equiv \frac{B^{+_n}_m-B^{-_n}_m}{2} = \frac{i^m}{4}\left(e^{2i\delta_m^{+_n}}-e^{2i\delta_m^{-_n}}\right).
 \end{align}

The wave function components $\varphi^{\gamma_n}(\mathbf{r})$ exhibit the asymptotic behaviors
\begin{align}
 \varphi^{a,n}(r,\theta) & \xrightarrow[r\rightarrow \,\infty]{} e^{ikr\cos\theta}+f^{a,n}(\theta)\;\frac{e^{ikr}}{\sqrt{r}}, \nonumber \\
 \varphi^{b,n+1}(r,\theta) & \xrightarrow[r\rightarrow \,\infty]{}
f^{b,n+1}(\theta)\;\frac{e^{ikr}}{\sqrt{r}},\label{asymptjb}
\end{align}
with the scattering amplitudes
\begin{equation}\label{ampliadd}
f^{\gamma_n}(\theta)=\sqrt{\frac{2}{\pi k}}\:\sum_{m=0}^{+\infty}\epsilon_m\cos(m\theta)e^{-i(m\frac{\pi}{2}+\frac{\pi}{4})}
 B^{\gamma_n}_m.
\end{equation}
These asymptotic behaviors highlight the initial state $|a\rangle$ of the incoming atoms since the wave function component $\varphi^{a,n}(r,\theta)$ is the only one to possess an incident plane wave $e^{i k x}$.

We then define the differential scattering lengths
\begin{equation}\label{dsl}
\lambda^{\gamma_n}(\theta)= |f^{\gamma_n}(\theta)|^2,
\end{equation}
as well as the total scattering lengths
\begin{equation}\label{tsl}
\lambda^{\gamma_n}
=\int_{-\pi}^{\pi}\lambda^{\gamma_n}(\theta)~d\theta,
\end{equation}
along with the \emph{dimensionless}  scattering lengths $\tilde{\lambda}^{\gamma_n}(\theta)$ and $\tilde{\lambda}^{\gamma_n}$~:
\begin{equation}
   \tilde{\lambda}^{\gamma_n}(\theta) \equiv \frac{\lambda^{\gamma_n}(\theta)}{2R}, \quad \tilde{\lambda}^{\gamma_n} \equiv \frac{\lambda^{\gamma_n}}{2R}.
\end{equation}

Using Eqs.~(\ref{Bab}) and (\ref{ampliadd}) the total scattering lengths $\lambda^{\gamma_n}$ can be expressed in terms of the phase shifts $\delta_m^{\pm_n}$. We get
\begin{equation}
\label{lambdagamman}
\lambda^{\gamma_n}=\frac{4}{k}\,\sum_{m=0}^{+\infty}\epsilon_m|B^{\gamma_n}_m|^2,
\end{equation}
with
\begin{equation}\label{tslabp}
\begin{aligned}
4|B^{a,n}_m|^2 = &
\sin^2(\delta^{+_n}_m-\delta^{-_n}_m) \\
& \, + 4\cos(\delta^{+_n}_m-\delta^{-_n}_m)\,\sin\delta^{+_n}_m\,\sin\delta^{-_n}_m,\\
4|B^{b,n+1}_m|^2 = & \sin^2(\delta^{+_n}_m-\delta^{-_n}_m).
\end{aligned}
\end{equation}

The differential scattering length $\lambda^{a,n}(\theta)$ [$\lambda^{b,n+1}(\theta)$] represents
the proportionality relation between the incoming flux of excited
atoms and the outgoing flux of these atoms scattered around the angle $\theta$ while remaining in their excited state $|a\rangle$ [while emitting a photon in the cavity and being de-excited in their lower state $|b\rangle$]. In short we will call hereafter $\lambda^{a,n}(\theta)$ [$\lambda^{b,n+1}(\theta)$] the \emph{no-deexcitation [photon-emission] differential scattering length}, and accordingly for the \emph{total scattering lengths} $\lambda^{a,n}$ and $\lambda^{b,n+1}$.

\subsection*{Transverse constant mode}

It is particularly instructive to consider the case of a transverse
constant mode
\begin{equation}\label{mesacyl}
v_{\mathrm{cyl}}(r)=\left\{
\begin{array}{ll}
1 & \hspace{10pt} r \leqslant R \vspace{5pt}\\
0 & \hspace{10pt} r > R
\end{array}\right.
\end{equation}
for which a fully analytical solution exists. Inside the cavity
($r\leqslant R$), the wave function components
$\varphi^{\pm_n}(\mathbf{r})$ are given by
\begin{equation}\label{psiBarrierVpm}
\begin{aligned}
        \varphi^{+_n}(r,\theta)={}& \sum_{m=0}^{+\infty}\epsilon_mA_m^{+_n}\cos(m\theta)\mathcal{I}_m(k^+_nr),\\
        \varphi^{-_n}(r,\theta)={}&\sum_{m=0}^{+\infty}\epsilon_mA_m^{-_n}\cos(m\theta)J_m(k^-_n r),
\end{aligned}
\end{equation}
where $k^{\pm}_n=\sqrt{|k^2 \mp \kappa_n^2|}$ represent the modified wave number of the atoms inside the cavity in the presence of the constant interacting potentials $\pm \hbar^2 \kappa_n^2/2M$, $A_m^{\pm_n}$ are coefficients determined similarly to the $B_m^{\pm_n}$ coefficients from the continuity conditions of the wave functions at the interface between the inside and the outside of the cavity, and
\begin{equation}
 \mathcal{I}_m(k^+_n r) = \left\{
\begin{array}{ll}
J_m(k^+_n r), &\hspace{10pt} k \geqslant \kappa_n,\vspace{8pt}\\
I_m(k^+_n r), &\hspace{10pt} k < \kappa_n,
\end{array}\right.
\end{equation}
with $I_m$ the modified Bessel function of the first kind of
order $m$.

The continuity conditions yield
\begin{equation}
\begin{aligned}
A_m^{+_n}={}&
\displaystyle\frac{-(2/\pi)i^{m+1}\epsilon_m}{k\,\mathcal{I}_m(k^+_n R)
{\H1m}'(k R)-k^+_n\mathcal{I}'_m(k^+_n R)\H1m(kR)},
\\
A_m^{-_n}={}&\displaystyle\frac{-(2/\pi)i^{m+1}\epsilon_m}{k\,J_m(k^-_n R)
{\H1m}'(kR)-k^-_n J'_m(k^-_n R)\H1m(kR)},
\end{aligned}
\end{equation}
and
\begin{equation}
\begin{aligned}
\frac{B_m^{+_n}}{A_m^{+_n}} & =\frac{i \pi}{2}\Big[k^+_n\mathcal{I}_m(kR)\mathcal{I}'_{m}(k^+_n R)
-k\,\mathcal{I}_m(k^+_n R)\mathcal{I}'_{m}(kR)\Big],
\\
\frac{B_m^{-_n}}{A_m^{-_n}} & =\frac{i \pi}{2}\Big[k^-_n J_m(kR)J'_{m}(k^-_n R)
-k\,J_m(k^-_n R)J'_{m}(kR)\Big],
\end{aligned}
\end{equation}
where the primes denote the derivatives of the functions with respect to their argument.

The phase shifts $\delta_m^{\pm_n}$ immediately follow from the relation [see Eq.~(\ref{Bps})]
\begin{equation}
\tan\delta_m^{\pm_n}=-\frac{iB_m^{\pm_n}}{i^m+B_m^{\pm_n}}.
\end{equation}

\section{Photon-emission and no-deexcitation scattering
lengths}\label{TSLSection}

In this section we investigate the photon-emission and no-deexcitation scattering lengths in the case of a transverse constant mode. We distinguish two regimes determined by the incident kinetic energy of the atom compared with the interaction energy $V^{+_n}$~: the high energy scattering or hot atom regime ($k \gg \kappa_n$) and the low energy scattering or cold atom regime ($k \ll \kappa_n$). In the cold atom regime, numerical results are also presented for a transverse gaussian mode.

\subsection{High energy scattering : hot atom regime}

In the hot atom regime, the phase shifts $\delta_m^{\pm_n}$ and the $B^{\pm_n}_m$ coefficients get negligible when the extension $r_m$ of the centrifugal barrier exceeds the inside region radius $R$ and prevents any ingoing cylindrical wave with wave number $k$ from being scattered by the potentials $V^{\pm_n}(r)$. According to Eq.~(\ref{relle}), this happens for $m \gtrsim m_{\ell}$ with
\begin{equation}
\label{melle}
m_{\ell} = \sqrt{(k R)^2 + 1/4}.
\end{equation}
This number yields the order of magnitude of the number of terms that is required in the sum~(\ref{lambdapmnsum}) for a good evaluation of the scattering lengths. It reads approximately $(k/\kappa) \kappa R$ and is thus significantly greater than the dimensionless \emph{interaction length} $\kappa R$.

\begin{figure}
\begin{center}
\includegraphics[width=.95\linewidth]{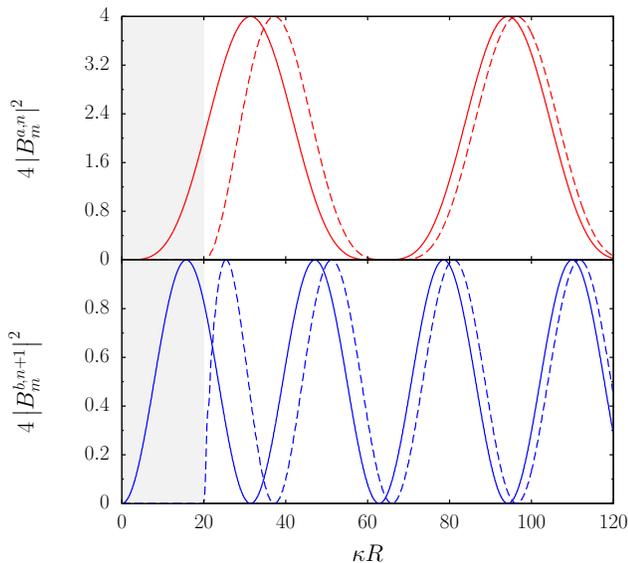}
\end{center}
\caption{(Color online) Plots of the $|B^{a,n}_m|^2$ (top) and $|B^{b,n+1}_m|^2$ (bottom) coefficients with respect to
the interaction length $\kappa R$ for a transverse constant mode and
the parameters $k/\kappa_n=10$ and $n=0$. Solid lines correspond to
$m=0$ and dashed lines to $m=200$. The grey shaded area is the
classically forbidden region for $m=200$ and $k/\kappa_n=10$.} \label{figlembhot}
\end{figure}

Figure~\ref{figlembhot} shows two coefficients $|B^{\gamma_n}_m|^2$ ($m=0$ and $m=200$) as a function of the interaction length $\kappa R$. As expected from the paragraph herein before, the coefficients $|B^{\gamma_n}_m|^2$ almost vanish for $\kappa R < \kappa r_m$. For $\kappa R \gg \kappa r_m$, the coefficients $|B^{\gamma_n}_m|^2$ display oscillations with respect to the interaction length that are well captured by the simple analytical expressions
\begin{equation}\label{rabioc}
4|B^{b,n+1}_m|^2\simeq\sin^2\left(\frac{\kappa_n R}{k/\kappa_n}\right)
\end{equation}
and
\begin{equation}\label{rabitl}
4|B^{a,n}_m|^2\simeq 4\sin^4\left(\frac{\kappa_n R}{2\,k/\kappa_n}\right).
\end{equation}

Equation~(\ref{rabioc}) is just like the Rabi formula for an atom
of velocity $v = \hbar k/M$ interacting resonantly with a single cavity mode during the time $\tau = 2R/v$. Similarly, Eq.~(\ref{rabitl}) is just like the
transition probability between two states of a three-level atom
coupled through a resonant two-photon transition during the time $\tau$~\cite{Yoo85}. This can be simply understood by noticing that the excited
scattered atoms are those atoms who have emitted a photon into and
subsequently absorbed another photon from the cavity field.

In the hot atom regime, the photon-emission differential scattering length $\ledb$
is given to an excellent approximation by (see the Appendix~\ref{AppendixA})
\begin{equation}\label{ledb}
    \ledb\simeq
\frac{(k/\kappa_n)^2}{4}\left[\frac{1-\sin\left(\frac{2\kappa_n R
\sqrt{(k/\kappa_n)^4\,\theta^2+1}}{k/\kappa_n}\right)}{\left[
    (k/\kappa_n)^4\,\theta^2+1\right]^{3/2}}\right],
\end{equation}
with $\theta$ ranging from $-\pi$ to $\pi$. Furthermore, when $\kappa_n R \gg k/\kappa_n$, the dimensionless photon-emission total scattering lengths $\tilde{\lambda}^{b,n+1}$ and $\tilde{\lambda}^{a,n}$ are very well approximated by
\begin{equation}\label{letemapprox}
    \tilde{\lambda}^{b,n+1}\simeq\frac{1}{2}\left[1-\frac{\pi}{2}J_0\left(\frac{2 \kappa_n
    R}{k/\kappa_n}\right)\right]
\end{equation}
and
\begin{equation}\label{letaapprox}
    \tilde{\lambda}^{a,n}\simeq\frac{1}{2}\left[3-2\pi J_0\left(\frac{\kappa_n
    R}{k/\kappa_n}\right)+\frac{\pi}{2}J_0\left(\frac{2\kappa_n
    R}{k/\kappa_n}\right)\right].
\end{equation}

\begin{figure}
\begin{center}
\includegraphics[width=.7\linewidth]{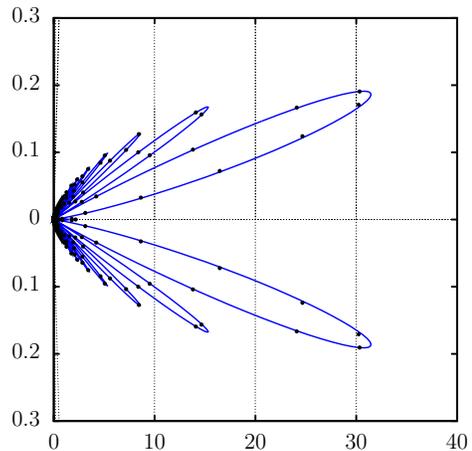}
\end{center}
\caption{(Color online) Polar plot of the differential scattering length $\ledb$
for $\kappa R=100$, $k/\kappa_n=10$ and $n=0$. Full dots correspond
to the approximated analytical formula (\ref{ledb}). 
Note the different scales on the $x$ and $y$ axes.} \label{figled}
\end{figure}

Figure~\ref{figled} shows a typical polar plot of $\ledb$ in the hot
atom regime. The atoms are only very slightly scattered from the incoming
direction with a spread in angle $\Delta\theta$ scaling like
$(k/\kappa_n)^{-2}$ [see Eq.~(\ref{ledb})]. The same conclusions hold for any other cavity mode functions, in particular the fundamental gaussian mode of a Fabry-Perot cavity (see the Appendix~\ref{AppendixA}). Similar scattering effects with hot atoms passing through a light
grating were experimentally reported in Refs.~\cite{Mos83,Gou86,Jur04}.

\begin{figure}
\begin{center}
\includegraphics[width=.95\linewidth]{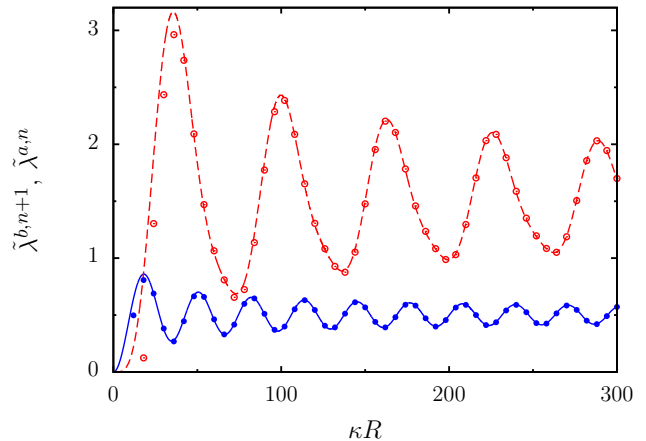}
\end{center}
\caption{(Color online) Total scattering lengths $\leta$ (top dashed line) and
$\letb$ (bottom solid line) with respect to the interaction length
$\kappa R$ for $k/\kappa_n=10$ and $n=0$. Full [empty] dots correspond
to the approximated analytical formula (\ref{letemapprox}) [(\ref{letaapprox})].}
\label{figlet}
\end{figure}

We compare in Fig.~\ref{figlet} the total scattering lengths $\leta$ and $\letb$ computed
from the exact Eq.~(\ref{tsl}) with those computed from
the approximated analytical formulas (\ref{letemapprox}) and
(\ref{letaapprox}). As can be observed in the figure, already for $\kappa_n R \gtrsim 50$ the agreement between the two
values is excellent.

We can gain some insight into Eq.~(\ref{letemapprox}) (and similarly
Eq.~(\ref{letaapprox})) by calculating the average emission
probability of a flux of excited atoms passing through the light field when considering their motion classically and ignoring any deflection of
their motion during their interaction with the cavity.
In that case, the interaction time $t_{\mathrm{int}}$ depends merely on the impact parameter $b=R\sin\theta$ according to
$t_{\mathrm{int}}=(2R\cos\theta)/v$ with $v$ the atomic velocity $\hbar k / M$. By integrating the Rabi emission probability over
all trajectories, we get
\begin{align}
\overline{\mathcal{P}}_{a\to b}
&{}=\frac{1}{\pi}\int_{-\pi/2}^{\pi/2}\sin^2\left(g\sqrt{n+1}\:
\frac{2R\cos\theta}{v}\right)d\theta\nonumber\\
\label{formmoyen}&{}=\frac{1}{2}\left[1-\,J_0\left(\frac{2\kappa_n
R}{k/\kappa_n}\right)\right],
\end{align}
since~\cite{Abr70}
\begin{equation}
    J_0(z)=\frac{1}{\pi}\int_{-\pi/2}^{\pi/2}\cos\big(z\cos(\theta)\big)\,d\theta.
\end{equation}

The resulting average emission probability [Eq.~(\ref{formmoyen})] differs from Eq.~(\ref{letemapprox}) only by the value of the coefficient in front of the Bessel function. The difference between these two expressions is the evidence that interference and deflection effects must be taken into account to give the proper result. Remarkably, Eq.~(\ref{formmoyen}) coincide with the photon-emission probability obtained by Saif~\etal~\cite{Sai01} in their study of a micromaser operating on the atomic scattering from a resonant standing wave in the Raman-Nath regime. This can be understood by noticing that in that regime, the atomic wavepacket is much larger than the periodicity of the standing wave, hence the atom-field interaction must be averaged over a full period. The same formula also appears in the work of Vaglica~\cite{Vag95} on Jaynes-Cummings model with atomic wavepackets.

\subsection{Low energy scattering : cold atom regime}

\begin{figure}
\begin{center}
\includegraphics[width=.95\linewidth]{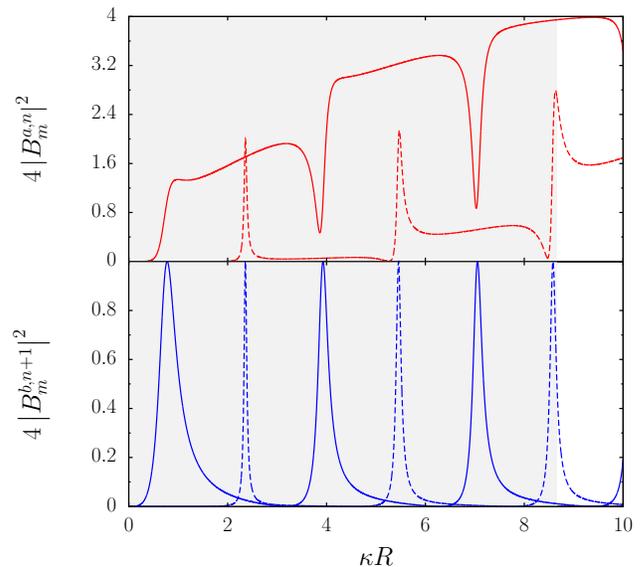}
\end{center}
\caption{(Color online) Plots of the $|B^{a,n}_m|^2$ (top) and $|B^{b,n+1}_m|^2$ (bottom) coefficients with respect to
the interaction length $\kappa R$ for $k/\kappa_n=0.1$ and $n=0$.
Solid lines correspond to $m=0$ and dashed lines to $m=1$. The grey shaded area is the classical forbidden region for $m = 1$.}
\label{figlemcold}
\end{figure}

In the cold atom regime, the scattering lengths exhibit a completely different behavior. The number $m_l$ of terms that contribute significantly to the scattering lengths in the sum~(\ref{lambdapmn}) is extremely limited.
We first show in Fig.~\ref{figlemcold} two coefficients $|B^{\gamma_n}_m|^2$ ($m = 0$ and $m = 1$) as a function
of the interaction length $\kappa R$. In contrast to the high energy
regime, fine resonances are here observed and those coefficients can differ largely from zero in the classically forbidden region. This is a signature of tunnelling effects of the atoms through the centrifugal barrier. When $\kappa_n R$ is much larger than $\kappa r_m$, the $|B^{b,n+1}_m|^2$ resonances are very well approximated by the simple analytical formulas
\begin{equation}\label{Bbmcoldanalyticeven}
    4|B^{b,n+1}_m|^2\simeq\frac{1-\cos \big(2\,\kappa_n R \sqrt{1+(k/\kappa_n)^2}\big)}{1+(k/\kappa_n)^{-2}\sin ^2\big(\kappa_n R \sqrt{1+(k/\kappa_n)^2}-\frac{\pi}{4}\big)},
\end{equation}
for even $m$, and
\begin{equation}\label{Bbmcoldanalyticodd}
    4|B^{b,n+1}_m|^2\simeq\frac{1+\cos \big(2\,\kappa_n R \sqrt{1+(k/\kappa_n)^2}\big)}{1+(k/\kappa_n)^{-2}\cos^2\big(\kappa_n R \sqrt{1+(k/\kappa_n)^2}-\frac{\pi}{4}\big)},
\end{equation}
for odd $m$. In both cases the resonance width is determined by the finesse
$(k/\kappa_n)^{-2}$ which increases as the atoms get colder and colder.

\begin{figure}
\begin{center}
\includegraphics[width=.75\linewidth]{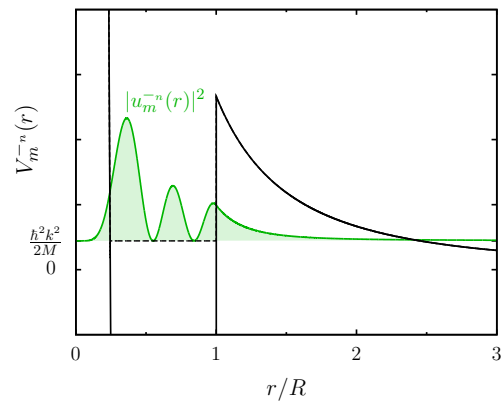}
\end{center}
\caption{(Color online) Effective potential $V^{-_n}_m(r)$ (black solid curve) felt by the radial wave function component $u_m^{-_n}(r)$ (green shaded curve) for $m = 3$, $k/\kappa_n=0.1$, $n=0$, and $\kappa R = 11.5287$. For that cavity size, the atomic kinetic energy matches a quasibound state energy of $V^{-_n}_m(r)$ (straight dashed line) and the stationary radial wave function is mainly located inside the potential well (tunnel effect), maximizing thereby the atom-field interaction and producing a resonance in the profile of $B^{b,n+1}_m$ and the related scattering lengths.}\label{quasibound}
\end{figure}

Very generally the position and width of
the resonances can be easily computed by noting that they are intimately linked to the quasibound states of the
effective potentials felt by the radial wave function components $u_m^{-_n}(r)$, i.e., the potentials
\begin{equation}\label{V}
V^{-_n}_m(r)=V^{-_n}(r)+\frac{\hbar^2}{2M}\frac{m^2-1/4}{r^2}.
\end{equation}
The quasibound states are those states related to unbound potentials containing a local minimum [like $V^{-_n}_m(r)$ for $m > 0$, see Fig.~\ref{quasibound}]. A particle initially confined in the potential well may remain there for an extremely long time, before escaping by tunnel effect to the lower potential region. A very similar situation is encountered in one-dimensional scattering of atoms by cavity fields subjected additionally to the action of gravity~\cite{Bas05a}.

The quasibound states are easily found by looking for the solutions of the stationary Schr\"odinger equation that represent a pure outgoing wave for $r \rightarrow \infty$~\cite{Lan65}. In the case of the transverse constant mode~(\ref{mesacyl}), these solutions are given by \begin{equation}\label{out}
u^{-_n}_m(r)=\left\{
\begin{array}{ll}
A_m\,J_m(k^-_n\,r), & \hspace{5pt} r\leqslant R, \vspace{5pt}\\
B_m\,H^{\scriptscriptstyle (1)}_m(kr), & \hspace{5pt} r>R. 
\end{array}\right.
\end{equation}
The continuity conditions of this
wave function and its first derivative at $r=R$ leads to
a system of equations whose secular equation reads
\begin{equation}\label{secular}
    k\,J_m(k^-_nR)\,{\H1m}'(kR)-k^-_n\,J'_m(k^-_nR)\,\H1m(kR)=0.
\end{equation}
The complex solutions of Eq.~(\ref{secular}) put in the form
\begin{equation}
    \kappa R_0-i\,\Gamma/2
\end{equation}
determine both the peak positions $\kappa R_0$ and
widths $\Gamma$ of the resonances of the $B^{-_n}_m$ coefficients, and thereby of the $B^{a,n}_m$ and $B^{b,n+1}_m$ coefficients.

\begin{figure}
\begin{center}
\includegraphics[width=.95\linewidth]{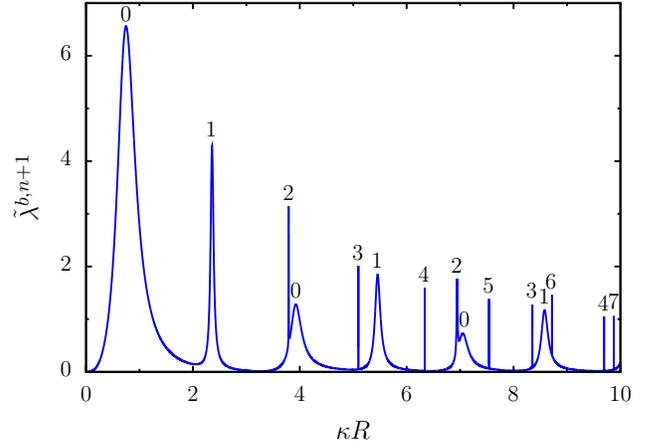}
\end{center}
\caption{(Color online) Dimensionless photon-emission total scattering length $\letb$ with respect to the interaction length
$\kappa R$ for $k/\kappa_n=0.1$ and $n=0$. Each resonance is labelled
by the integer $m$ of the $B^{b,n+1}_m$ coefficient it stems from.} \label{figletcold}
\end{figure}

Figure~\ref{figletcold} shows the dimensionless photon-emission total scattering length $\letb$ as a function of the interaction length
$\kappa R$. As $\letb$ is the weighted sum of all $|B^{b,n+1}_m|^2$ coefficients [see Eq.~(\ref{lambdagamman})], all resonances of these coefficients add up and generate the ensemble of resonances that are observed in Fig.~\ref{figletcold}. This allows us to label these resonances by the integers $m$ of the $B^{b,n+1}_m$ coefficients they originate.
The $m = 0$ resonances are the only ones not to be linked to quasibound states since $V_0^{-_n}(r)$ is strictly attractive and does not contain any local minimum. They are the signature of a low energy scattering process upon a purely attractive potential and this explains why they are significantly broader.

\begin{figure}
\begin{center}
\includegraphics[width=.95\linewidth]{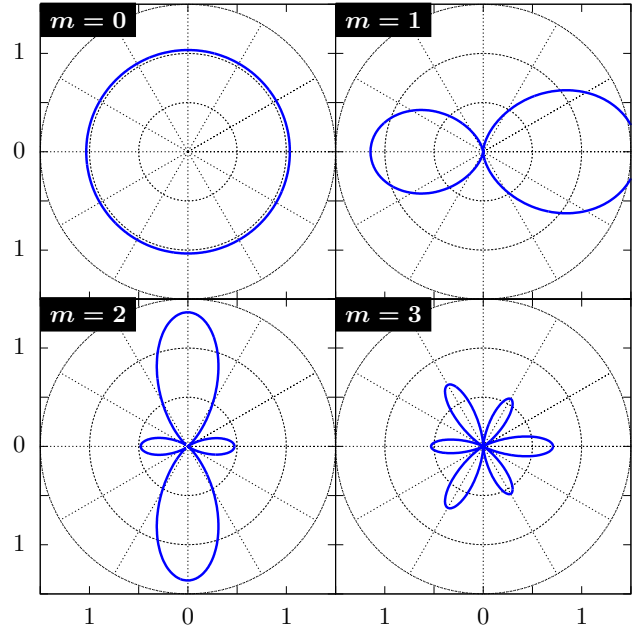}
\end{center}
\caption{(Color online) Dimensionless photon-emission differential scattering length $\ledb$ for $k/\kappa_n=0.1$, $n=0$ and four of the
five first resonances observed in Fig.~\ref{figletcold} : $\kappa R=0.72890$
($m=0$), $\kappa R=2.35741$ ($m=1$), $\kappa R=3.79243$ ($m=2$), and
$\kappa R=5.09697$ ($m=3$).} \label{figledcold}
\end{figure}

When the resonances do not overlap, the differential scattering length $\ledb$ identifies to the squared absolute value of a single term of Eq.~(\ref{ampliadd}) and we have in excellent approximation
\begin{equation}\label{ledbcold}
    \ledb \propto \frac{\cos^2(m\theta)}{kR}.
\end{equation}
We display accordingly in Fig.~\ref{figledcold} the scattering patterns $\ledb$ for four different interaction lengths corresponding to four resonances
 of Fig.~\ref{figletcold}. In contrast to the high energy scattering
regime, cold atoms can be backscattered with a high probability.
Also by varying the interaction length $\kappa R$, the scattering
pattern can be tuned from a resonance pattern to another.

\begin{figure}
\begin{center}
\includegraphics[width=.95\linewidth]{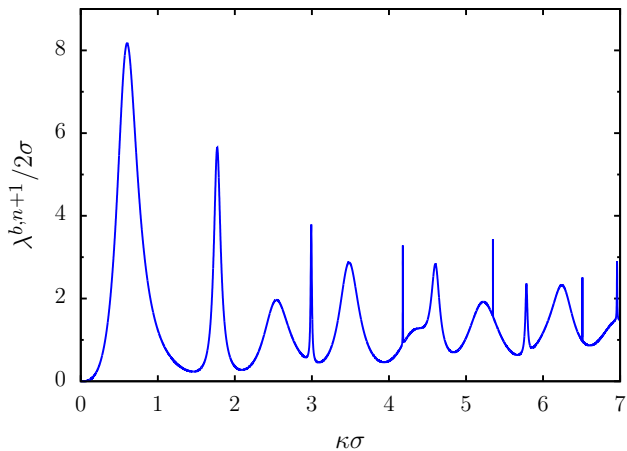}
\end{center}
\caption{(Color online) Dimensionless photon-emission total scattering length $\lambda^{b,n+1}/2\sigma$ with respect to the interaction length $\kappa \sigma$ for a gaussian mode and for $k/\kappa_n=0.1$ and $n=0$.}
\label{figletcoldgauss}
\end{figure}

Finally we consider the case of a gaussian mode of standard deviation $\sigma$ [$v_{\mathrm{gauss}}(r)=\exp(-r^2/2 \sigma^2)$]. We show in Fig.~\ref{figletcoldgauss} a plot of the photon-emission total scattering length $\lambda^{b,n+1}/2\sigma$ with respect to the \emph{gaussian mode interaction length} $\kappa \sigma$. The results are qualitatively the same as those obtained for the transverse constant mode. This is not surprising since the underlying physical mechanisms are identical~: the attractive part of the gaussian potentials formed by the light field in combination with the centrifugal barriers exhibits quasibound states giving rise to resonances in the scattering length. Compared with the transverse constant mode, the resonances are here broader and overlap more significantly.
For well chosen interaction lengths, the differential scattering length displays similar patterns as those of Fig.~\ref{figledcold}.

\section{Conclusions}
\label{SummarySection}
In this paper, we have presented the quantum theory of the cold atom scattering by cavity fields in a two-dimensional geometry in a distinct regime from the usual Raman-Nath, Bragg and Stern-Gerlach regimes. General expressions for the photon-emission and no-deexcitation scattering lengths have been derived. The connection with the classical Rabi limit in the hot atom regime has been provided and approximated analytical results have been given following a semiclassical approach. In the cold atom regime, we have highlighted the very peculiar properties of the scattering process and their interpretation in terms of quasibound states of the atomic motion induced by the attractive potential of the cavity light field in combination with the centrifugal barriers. Realistic gaussian mode functions have been finally discussed.

\begin{acknowledgments}
This work has been supported by the Belgian
Institut Interuniversitaire des Sciences Nucl\'eaires (IISN). J.M.\
thanks the Belgian F.R.S.-FNRS for financial support. T.B. thanks E.\ Solano and N. Zagury for fruitful discussions and hospitality at Universidade Federal do Rio de Janeiro (Brazil). 
\end{acknowledgments}

\appendix*

\section{}
\label{AppendixA}

In this Appendix, Eqs.~(\ref{ledb})-(\ref{letaapprox}) are derived using a
semiclassical approach. In the hot atom regime, the number $m_l$ of Eq.~(\ref{melle}) is large and the sum over
$m$ in Eq.~(\ref{ampliadd}) can be replaced by an integral. The scattering amplitudes can then be approximated using the
eikonal approximation~\cite{Adh08} by
\begin{equation}
    f^{\pm_n}_{\mathrm{eik}}(\theta)=-\frac{ik}{\sqrt{2\pi}}\int_0^{+\infty}
    \cos(kb\theta)\left(e^{2i\delta^{\pm_n}_{\mathrm{eik}}}-1\right)db,
\end{equation}
where $m=k b$, $b$ is the impact parameter, and the
eikonal phase shifts $\delta^{\pm_n}_{\mathrm{eik}}$ are given by
\begin{equation}
    \delta^{\pm_n}_{\mathrm{eik}}=-\frac{1}{2k}\int_b^{+\infty}\frac{\pm \kappa_n^2 v(r)\, r \,
    dr}{\sqrt{r^2-b^2}},
\end{equation}
with $v(r)$ the considered mode function. For a transverse constant mode, we have for $R>b$
\begin{equation}
    \delta^{\pm_n}_{\mathrm{eik}}=\mp \frac{\kappa_n^2}{2k}\sqrt{R^2-b^2}.
\end{equation}
It follows the photon-emission scattering amplitude 
\begin{equation}\label{ampbeik}
    f^{b,n+1}_{\mathrm{eik}}(\theta)=-\sqrt{\frac{2k}{\pi}}\int_0^{+\infty}\cos(kb\theta)\sin\left(\frac{\kappa^2_n}{k}\sqrt{R^2-b^2}\right)db,
\end{equation}
and, using the stationary phase method~\cite{Ble86} to evaluate the integral, the photon-emission differential scattering length (in terms of dimensionless parameters)
\begin{equation}
    \tilde{\lambda}^{b,n+1}_{\mathrm{eik}}(\theta) \simeq \frac{(k/\kappa_n)^2}{4}\left[\frac{1-\sin\left(\frac{2\kappa_n R
\sqrt{(k/\kappa_n)^4\,\theta^2+1}}{k/\kappa_n}\right)}{\left[
    (k/\kappa_n)^4\,\theta^2+1\right]^{3/2}}\right].
\end{equation}

A second use of the stationary phase method for integrating the differential scattering length $\tilde{\lambda}^{b,n+1}_{\mathrm{eik}}(\theta)$ over all angles yields for $\kappa_n R\gg 1$
\begin{equation}
    \tilde{\lambda}^{b,n+1}_{\mathrm{eik}}\simeq\frac{1}{2}-\sqrt{\frac{\pi}{32} \frac{k/\kappa_n}{\kappa_n
    R}}\cos\left(\frac{2 \kappa_n
    R}{k/\kappa_n}-\frac{\pi}{4}\right),
\end{equation}
that simplifies further when $\kappa_n R\gg k/\kappa_n$ into
\begin{equation}
    \tilde{\lambda}^{b,n+1}_{\mathrm{eik}}\simeq\frac{1}{2}\left[1-\frac{\pi}{2}J_0\left(\frac{2 \kappa_n
    R}{k/\kappa_n}\right)\right].
\end{equation}

Proceeding along the same line allows one to obtain the no-deexcitation total scattering length
\begin{equation}
    \tilde{\lambda}^{a,n}_{\mathrm{eik}}\simeq\frac{1}{2}\left[3-2\pi J_0\left(\frac{\kappa_n
    R}{k/\kappa_n}\right)+\frac{\pi}{2}J_0\left(\frac{2\kappa_n
    R}{k/\kappa_n}\right)\right].
\end{equation}

In the case of a gaussian mode of standard deviation $\sigma$ [$v(r)=\exp(-r^2/2\sigma^2)$], we have
\begin{equation}
    \delta^{\pm_n}_{\mathrm{eik}}=\mp
    \frac{\kappa_n^2\sigma}{2k}\sqrt{\frac{\pi}{2}}e^{-b^2/2\sigma^2}
\end{equation}
and the photon-emission scattering amplitude reads
\begin{align}\label{gausseik}
    f^{b,n+1}_{\mathrm{eik}}(\theta) & =
    -\sqrt{\frac{2k}{\pi}} \times \nonumber \\
    &\!\!\!\!\!\!\!\!\!\!\!\! \int_0^{+\infty}
    \cos(kb\theta)\sin\left(\frac{\kappa^2_n\sigma}{k}\sqrt{\frac{\pi}{2}}e^{-b^2/2\sigma^2}\right)db.
\end{align}

Writing the integrand of (\ref{gausseik}) as a sum of exponentials
and calculating the stationary points $b_s$ of the various phases yields
\begin{equation}
b_s=\pm i \sigma \sqrt{W\left(-2k^4\theta^2/\pi\kappa_n^4\right)},
\end{equation}
where $W$ is the Lambert $W$ function~\footnote{The Lambert $W$ function, also called Omega function, is the inverse of the function $f(w) = w e^w$, i.e., the function $W(z)$ verifying $W(z)e^{W(z)} = z$.}. It follows
that the stationary points $b_s$ are real only if
\begin{equation}
W\left(-\frac{2}{\pi}\left(\frac{k}{\kappa_n}\right)^4\theta^2\right)\leq 0.
\end{equation}
This
condition translates into the angle $\theta$ as
\begin{equation}\label{thetacgauss}
\theta\leq \theta_c\equiv\sqrt{\frac{\pi}{2e}}\,\left(\frac{k}{\kappa_n}\right)^{-2}.
\end{equation}
From a physical point of view, this means that in the hot atom regime
and for $\kappa_n \sigma\gg 1$ the photon-emission differential
scattering length $\lambda^{b,n+1}(\theta)$ drops rapidly down to zero with $\theta$ when $\theta\gtrsim\theta_c$.
It is interesting to notice that the critical angle $\theta_c$ only depends on the amplitude of the scattering potential (through the term $\kappa_n$) and not on the extension of this potential.

\end{document}